\documentclass[12pt,preprint]{aastex}

\shortauthors{Gizis et al.}
\shorttitle{Distance to 2M1207-39}

\begin{document}

\title{The Trigonometric Parallax of the Brown Dwarf Planetary System 2MASSW J1207334-393254}

\author{John E. Gizis\altaffilmark{1}}
\affil{Department of Physics and Astronomy, University of Delaware, Newark, DE 19716, USA; gizis@udel.edu}
\author{Wei-Chun Jao\altaffilmark{1}, John P. Subasavage\altaffilmark{1}, Todd J. Henry\altaffilmark{1}}
\affil{Department of Physics and Astronomy, Georgia State University, Atlanta, GA 30302-4106, USA; jao@chara.gsu.edu, subasavage@chara.gsu.edu, thenry@chara.gsu.edu}

\altaffiltext{1}{Based on observations obtained at Cerro Tololo Inter-American Observatory, a division of the National Optical Astronomy Observatories, which is operated by the Association of Universities for Research in Astronomy, Inc. under cooperative agreement with the National Science Foundation.}

\begin{abstract}
We have measured a trigonometric parallax to the young brown dwarf 2MASSW J1207334-393254.  The distance ($54.0^{+3.2}_{-2.8} ~\rm{pc}$) and space motion confirm membership in the TW Hydrae Association.  The primary is a $\sim 25 M_{jup}$ brown dwarf.  We discuss the "planetary mass" secondary, which is certainly below the deuterium-burning limit but whose colors and absolute magnitudes pose challenges to our current understanding of planetary-mass objects.\end{abstract}


\section{Introduction}

The M8 brown dwarf 2MASSW J1207334-393254 (hereafter 2M1207A) is proving to be an important system for studying the formation of substellar objects.  It was discovered by \citet{gizis2002} in a search for brown dwarf members of the $\sim 10$ Myr old TW Hydrae Assocation \citep{twa}.  2M1207A is a very-low-mass substellar analog to a classical T Tauri star: It has broad, variable H$\alpha$ emission due to accretion \citep{mohantyvrad,broadha}, mid-infrared excess due to a disk \citep{gemini,spitzer}, ultraviolet emission due to hot accreted gas and warm circumstellar molecular hydrogen gas \citep{uv}, and forbidden oxygen emission due to an outflow\citep{outflow}.  Despite its youth, it is not detected in X rays \citep{xrays} or radio \citep{radio}, so is apparently relatively magnetically inactive.

\citet{planet1} discovered a red companion (2M1207B), 5 magnitudes fainter in the K band.  Common proper motion confirms that this is a bound pair \citep{planet2,song} with a separation of $773 \pm 1.4 {\rm mas}$.  The secondary has a late-L spectral type \citep{mohantydisk}.  The inferred luminosity implies a mass $\sim 5 M_J$ \citep{planet1, song}, although \citet{mohantydisk} suggest that the secondary is  $8\pm2$ jupiter masses and viewed through an edge-on disk.   

Because the TWA is a relatively nearby, loose association there has been some confusion on the distance to the system.  \citet{planet1} adopted a distance of 70 pc, on the basis of theoretical models of brown dwarf evolution.  The Hipparcos distance of TW Hya itself is $56^{+8}_{-6} {\rm pc}$ \citep{hipparcos}.  \citet{mamajek} used the moving cluster distance method to estimate the distance to 2M1207A to be $53 \pm 7$ pc, while \citet{song} used the same method, but an updated proper motion and a different group membership list to estimate $59 \pm 7$ pc.   
With uncertainties in the distance to the TW Hya group of   $\sim 15\%$, firm conclusions about the natures of 2M1207 A and B, as  well as other members of the group, have been elusive."  Here we present the first trigonometric parallax for 2M1207A.  We confirm that it is a member of the TW Hya Association and put put constraints on the planet candidate 2M1207B.

\section{Parallax Results and Dicussion}

Observations of 2M1207A in the $I_{KC}$ band were obtained at the CTIO 0.9m telescope by the RECONS group via the SMARTS Consortium.  There are 54 parallax frames obtained over 2.14 years.  The observing techniques and data reduction are fully described by \citet{jao}.   The resulting relative parallax is $\pi_{rel} = 17.93 \pm 1.03~{\rm mas}$.   $VRI$ photometry was obtained in July 2007 on five nights using the same telescope and reduced as described in \citet{jao}.   We estimate the correction to absolute parallax to be $0.58 \pm 0.05~{\rm mas}$ on the basis of photometry of the seven reference stars (Table 1.)  The absolute parallax is therefore $18.51 \pm 1.03 ~{\rm mas}$, for a distance of $54.0^{+3.2}_{-2.8} \rm{pc}$.  The observed proper motion is $66.7 \pm 1.5 ~{\rm mas yr}^{-1}$ at position angle $\theta=250.0 \pm 2.4$ degrees.  

The distance and proper motion of 2M1207 is consistent with TWA membership.  The position angle expected for motion towards \citet{mamajek}'s TWA convergent point is 251.4 degrees, consistent with the measured proper motion.  Using \citet{mohantyvrad}'s radial velocity of $+11.2 \pm 2.0 ~{\rm km~s}^{-1}$ for 2M1207A, the (U,V,W) space velocities are ($-8, -18, -4$) ${\rm km s}^{-1}$, consistent with Mamajek's centroid group value of ($-10.2, - 17.1, - 5.1$) ${\rm km s}^{-1}$ .    In particular, the measured distance rules out any association with the background Lower Centaurus Crux discussed by \citet{mamajek}.    Using \citet{song}'s measurements and our distance, the projected separation is $41.7 \pm 2.3$ A.U.   

{\it The Primary and its Disk:}   \citet{mohantydisk} found 2M1207A to be $24 \pm 6 M_{J}$ brown dwarf. Because they used Mamajek's value of 53 pc as the distance, this mass is not changed significantly by a distance increase of 2\%:   2M1207A is best understood as a $\sim 25 M_J$ brown dwarf.  The disk parameters derived by \citet{riaz} also remain unchanged because they used the same distance.   The observed $V-K_s = 8.00 \pm 0.19$ is consistent with the M8 spectral type and suggests the accretion rate at the time was $\lesssim 10^{-11} M_\odot {\rm yr}^{-1}$ (see Figure 4 of Riaz \& Gizis 2007.)   In Figure~1, we plot the H-R diagram of the local field population and 2M1207A.  Like the young M dwarf AU Mic (Gl 803), 2M1207A lies $\sim 1.5$ magnitudes above the main sequence in the $M_V$ vs $V-K$ diagram, confirming youth.

 {\it The Secondary:}    The usual procedure for analyzing 2M1207B is to assume a bolometric correction appropriate to late-L dwarfs, and then fit the luminosity to evolutionary models.  \citet{planet1} estimated $5 \pm 2 M_J$ for 70 pc, \citet{song} estimated $5 \pm 3 M_J$ for 59 pc, and \citet{mamajek} estimated $3-4 M_J$ for 53 pc.   The trigonometric parallax would therefore support the last two estimates.  \citet{mohantydisk}, however, noted an inconsistency with this procedure. They argued that their H and K-band near-infrared spectra of 2M1207B were best fit by an effective temperature of $1600 \pm 100$K.  However, for the $3-5 M_J$ fits, the expected effective temperature is more like $1000-1200$ K.  They suggest the best resolution is that 2M1207B is viewed through an edge-on gray disk, and that therefore it is more luminous than otherwise estimated.  2M1207B is then a $8 \pm 2M_J$ planetary mass brown dwarf.   The wide separation and mass ratio ($q \approx 0.2-0.3$) suggests this planetary-mass object did not form through core accretion \citep{planet2}.
 
Without rejecting the possibility of a edge-on disk, we argue that available evidence does not rule out a low temperature for 2M1207B.  In Figure~2, we plot colors and absolute magnitudes for late-M, L and T dwarfs with parallaxes \citep{hipparcos, dahn, tinney, vrba, henry06}.  All the previous attempts to fit 2M1207B noted that it is red compared to field brown dwarfs, which can be attributed to having more dust in the photosphere.  The faintness at J-band measured by  \citet{mohantydisk} ($\Delta J = 7.0\pm0.2$) is supported by the NICMOS F110M measurement of \citet{song} ($\Delta m_{110M} =  7.17 \pm 0.15$).    \citet{planet1} measured $\Delta K=4.98$ and $K = 16.93 \pm 0.11$ for 2M1207B.  On the other hand, it is clear from Mohanty et al.'s (2007) ($SN \approx 3-10$) spectrum that there is deep water absorption but little methane absorption.   We conclude that 2M1207B has a spectral energy distribution that is L-type, but very red.   Reversing the usual procedure, in Figure~3 we plot the required K-band bolometric correction required to fit the \citet{chabrier} models at an age of 10 Myr.    Observed bolometric corrections for field L and T dwarfs from \citet{golimowski} as a function of temperature for an assumed age of $3 Gyr$ are also shown.    The L to T transition is believed to occur at $\sim 1300$K, and can be marked in Figure 3 by the change in bolometric corrections.   The transition is related to the change in dust properties in photosphere (see Kirkpatrick 2005 for a review), and some of the first fits to  late-L dwarf spectra gave incorrect values of $\gtrsim 1600$K due to the failure of cooler models to resemble the real spectra.   This simply reflects the extreme difficulty of modelling the temperature range 1200-1400K, and indeed In light of this, no models succeed in fully explaining both the blue hook and brightening in J-band of field T dwarfs.   Analysis of the luminosity (see Kirkpatrick) has been the most reliable way to derive temperatures.  This history suggests to us that the existing fit must be viewed with caution --- while apparently very good, it is inconsistent with the absolute magnitudes unless an edge-on disk is invoked.  Although Mohanty et al. show that the DUSTY models do fit the observed color of 2M1207B, it must be noted that the same models predict very red colors for field L dwarfs, which are not observed.  Indeed, \citet{chabrier} note that "the DUSTY and COND models represent extreme situations which bracket the more likely intermediate case resulting from complex, and presently not understood, thermochemical and dynamical processes."  We think it plausible that existing DUSTY models fail to properly model the dust in low surface gravity dwarfs, and that 2M1207B might therefore be $\sim 1200$K, as expected for the $\sim 5 M_{jup}$ model.   As an example how this might occur, we note that \citet{tsuji} invokes a parameter, $T_{cr}$, that characterizes the thickness of the clouds, and argues that the wide range of colors for field objects near $1400$K is due to changes in this parameter in otherwise similar brown dwarfs.  In one case, \citet{tsuji} is able to fit an L6.5 dwarf with $T_{eff} = 1700$K or $T_{eff}=1300$K  (without methane absorption) by varying $T_{cr}$ by only 100K.  Evidently an extremely red color like 2M1207B could be obtained for a low $T_{eff}$ with $T_{cr} < 1700$K -- that is, a very thick cloud compared to field L  dwarfs.  This would be the opposite situation from field T dwarfs, where the cloud becomes thinner ($T_{cr}$ increases.)  Similarly, in the \citet{frain} models, a parameter, $f_{rain}$, represents sedimentation, and redder colors are produced by smaller values of $f_{rain}$ (i.e., less precipitation and thicker clouds.)  Regardless of how the proper degree of dust is produced, if our speculation that $T_{eff} \approx 1200$K is correct, the implied $BC_K \approx 2.5$ would require that more of 2M1207B's energy is escaping at wavelengths longward of 3 microns than in field L dwarfs.  In any case, the observed colors and spectrum do not match any field brown dwarf, so there is not much doubt that atmosphere is dustier, but a low temperature remains speculative.  

Unfortunately, there is a third problem with estimating the mass of 2M1207B.   \citet{marley} have investigated the dependence of the structure models to the initial conditions, and found that the luminosity is very sensitive to the initial conditions for up to 100 Myr.  In specifically discussing the case of 2M1207B, they note that  for  a "warm start" rather than usually assumed "hot start," the best fit mass is  $8 M_{jup}$ rather than $5 M_{jup}$.   The situation, therefore, is that given the now known distance, 2M1207B may be $\sim 5 \pm 2 M_{jup}$ {\it if} current structural models are correct, the red color implies a cool temperature, and there is no disk, but both \citet{mohantydisk} and \citet{marley} present plausible scenarios in which the mass is higher.  
 
\section{Conclusions}

We have measured the trigonometric parallax of 2M1207 and found that the distance and space motion are consistent, as expected, with membership in the TW Hydrae Association.  Indeed, 2M1207 now has a more precise distance determination than TW Hya itself.  There are no difficulties in modelling the primary:  It is a $\sim 25 M_{jup}$   mass brown dwarf that is accreting from a circumstellar disk.  The faint secondary remains problematic.  Because we do not know the appropriate initial conditions, do not have a model atmosphere that reproduces the colors, and do not know whether or not it is observed through a disk, a case can be made that 2M1207B's mass is as low as $3 M_J$ or as high as $8 M_J$.  Our best estimate is $\sim 5 M_{jup}$ if 2M1207B is not viewed through a disk.    Further study of this planetary mass object is needed; we particularly need to know if it has an effective temperature of $1600$K , $1200$K, or even less.  

\acknowledgments

We thank Eric Mamajek and Davy Kirkpatrick for useful discussions.  We thank Charlie Finch for his initial reduction of the parallax data.  We thank the anonymous referee for a discussion of the Tsuji paper.  We also thank the members of the SMARTS Consortium, without whom the parallax observations could not have been made.  Support for this work was provided by NASA Research Grant NNG06GJ03G.  The RECONS parallax program has been supported by the NASA/NSF NStars Project, NASA's Space Interferometry Mission (SIM), the National Science Foundation (grant AST 05-07711), and Georgia State University.  Research has benefitted from the M, L, and T dwarf compendium housed at DwarfArchives.org and maintained by Chris Gelino, Davy Kirkpatrick, and Adam Burgasser.

\begin{deluxetable}{rrr}
\tablewidth{0pc}
\tablenum{1}
\tablecaption{2M1207}
\tablehead{
\colhead{Item} &
\colhead{2M1207A} &
\colhead{2M1207B} 
}
\startdata
$V$ & $19.95 \pm 0.19$ & \nodata\\
$R $ & $17.99 \pm  0.07$ & \nodata\\
$I$ & $15.92 \pm 0.05$ & \nodata\\ 
$J$ & $13.00 \pm 0.03$ &  $20.0 \pm 0.2$ \\
$H$ & $12.39 \pm 0.03$ & $18.09 \pm 0.21$ \\
$K$ & $11.95 \pm 0.03$ &  $16.93 \pm 0.11$ \\
$\pi_{rel}$ (mas)& $17.93 \pm  1.03$ \\
$\pi_{ref}$ (mas)& $0.58 \pm  0.05$ \\
$\pi_{abs}$ (mas)& $18.51 \pm  1.03$ \\
$M_V$ & $16.29 \pm 0.22$ & \nodata\\
$M_J$  & $9.34 \pm 0.12$ & $16.3 \pm 0.3$ \\ 
$M_H$ & $8.73 \pm 0.12$ & $14.4 \pm 0.3$ \\
$M_K$ & $8.29 \pm 0.12$ & $13.27 \pm 0.16$\\
Mass ($M_{jup}$) & $\sim 25$ & $\sim 3-8$ \\ 
\enddata
\tablecomments{VRI and astrometry from this paper; JHK for 2M1207A from 2MASS \citet{skrutskie}; J for 2M1207B from \citet{mohantydisk}; H and K from \citet{planet1}}
\end{deluxetable}


\begin{figure}
\epsscale{0.7}
\plotone{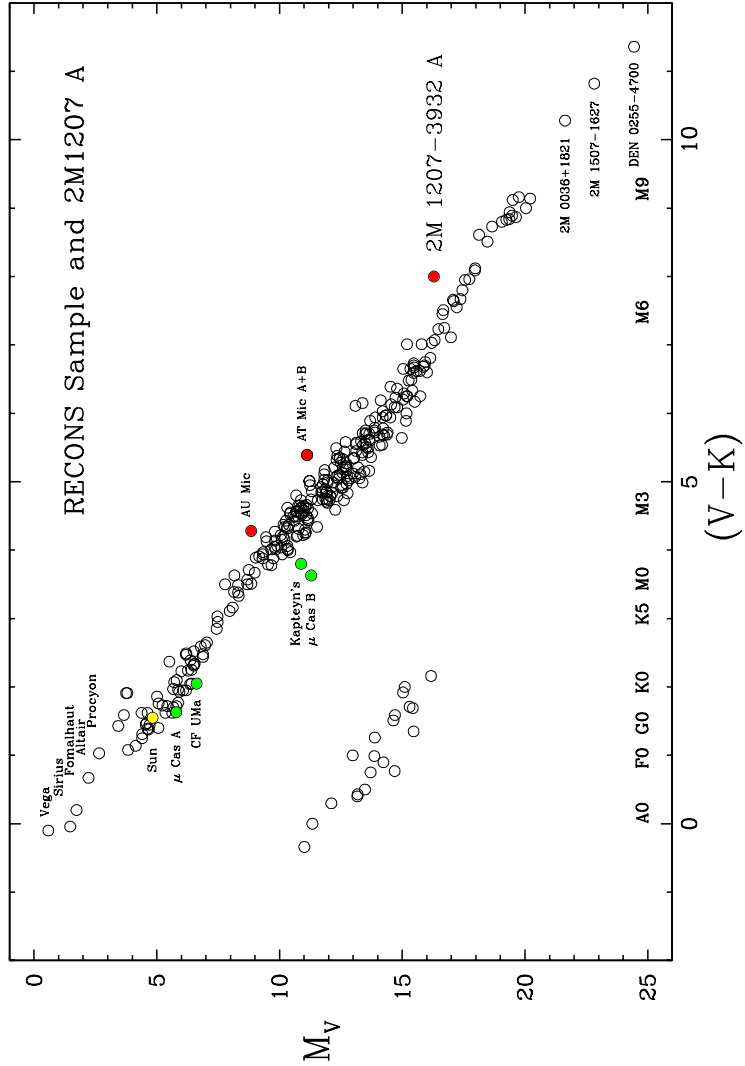}
\caption{2M1207A is plotted among the known members of the RECONS 10 pc sample \citep{henry06}. Highlighted points include the Sun, the three subdwarf systems $\mu$ Cas AB, CF UMa, and Kapteyn's Star, and the triple system AU Mic/AT Mic AB. The latter system is estimated to be $\sim 12$ Myr in age \citet{betapic}, causing the points to be elevated above the main sequence, as expected (the point for AT Mic AB actually represents each component, not the combined light of both components --- the two stars are nearly identical in brightness). Because of youth, 2M1207 A is also elevated above the main sequence, much like its nearer counterparts, AU Mic and AT Mic.
\label{fig1}}
\end{figure}

\begin{figure}
\plotone{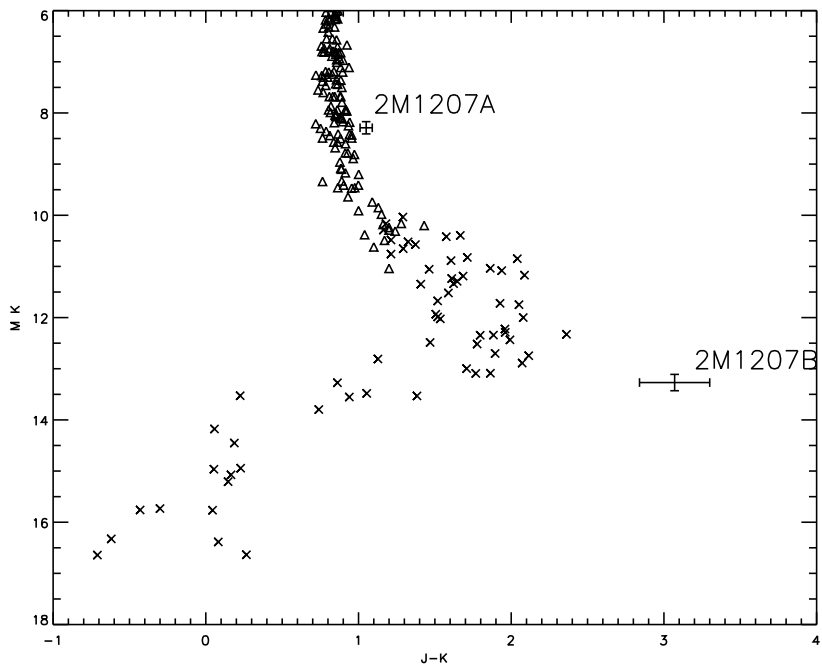}
\caption{H-R diagram showing nearby field M dwarfs (triangles), field L and T dwarfs (crosses), and 2M1207A and B (points with error bars.)  While 2M1207A is simply an overluminous M8 with low surface gravity as expected for a very young brown dwarf, the color of 2M1207B is much redder than field objects.\label{fig2}}
\end{figure}

\begin{figure}
\plotone{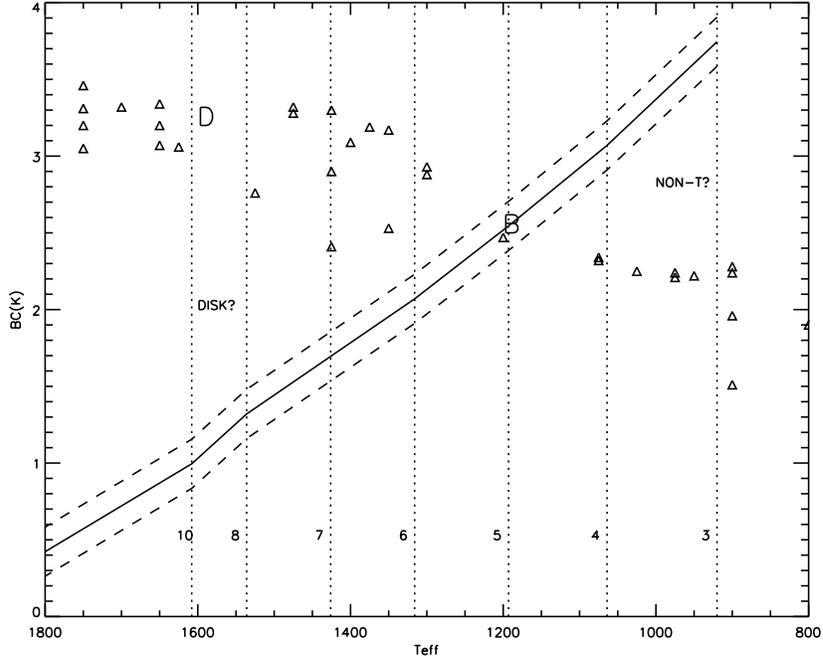}
\caption{The K-band bolometric correction (solid line) required to match the observed $M_K$ to the predicted luminosity of the 10 Myr \citet{chabrier} models as a function of the model temperature.     The dashed lines above and below the solid line shows the uncertainties due to the K magnitude and parallax uncertainties, but do not account for age or model uncertainties.  Also shown are measured bolometric corrections for old field L and T dwarfs by \citet{golimowski}.  The temperature predicted by different masses (in $M_{jup} = 0.001 M_\odot$) of the models is marked by the dotted lines.  If the \citet{chabrier} models are correct, 2M1207 could lie anywhere along the solid line --- on the other hand, if the spectral energy distributions of young objects are similar to field brown dwarfs, then 2M1207B should lie near the region populated by open triangles.  We draw attention to the point ($1200$K, 2.5, marked by a B), which is a plausible solution that matches the $5 M_{jup}$ model.  We also draw attention to the point ($1600$K, 3.2, marked by a D) which corresponds to the \citet{mohantydisk} solution of an $8 M_{jup}$ model --- in this case, extinction due to a disk  explains the difference between D and the model ($1600$K, 1.0)  We note that the 3 jupiter mass would require a bolometric correction of  3.7 at a temperature where field T dwarfs have  $BC_K\approx 2$ \label{fig3}}
\end{figure}


\begin{thebibliography}{}

\bibitem[Chauvin et al.(2004)]{planet1} Chauvin, G., et al. 2004, \aap, 425, L29

\bibitem[Chauvin et al.(2005)]{planet2} Chauvin, G., et al. 2005, \aap, 438, L25

\bibitem[Chabrier et al.(2000)]{chabrier} Chabrier, G., Baraffe, I., Allard, F., \& Hauschildt, P. 2000, \apj, 542, 464

\bibitem[Dahn et al.(2002)]{dahn} Dahn, C.C., et al. 2002, \aj, 124, 1170

\bibitem[Gizis(2002)]{gizis2002} Gizis, J.E., 2002, \apj, 575, 484

\bibitem[Gizis \& Bharat(2004)]{xrays} Gizis, J.E., \& Bharat, R.  2004, \apj, 608, L113

\bibitem[Gizis et al.(2005)]{uv} Gizis, J.E., Shipman, H.L., \& Harvin, J.A. 2005, ApJ, 630, L89

\bibitem[Golimowski et al.(2004)]{golimowski} Golimowski, D.A., et al.  2004, \aj, 127, 3516

\bibitem[Henry et al.(2006)]{henry06} Henry, T.J., Jao, W.-C., Subasavage, J.P., Beaulieu, T., Ianna, P.A., 
Costa, E., \& Mendez, R.A.  2006, \aj, 132, 2360

\bibitem[Jao et al.(2005)]{jao} Jao, W.-C., et al.  2005, \aj, 129, 1954

\bibitem[Jayawardhana et al.(2003)]{jaya} Jayawardhana, R., Ardila, D. R., Stelzer, B., \& Haisch, K.E. 2003, \aj, 126, 1515

\bibitem[Kirkpatrick(2005)]{kirk} Kirkpatrick, J.D.  2005, \araa, 43, 195

\bibitem[Mamajek(2005)]{mamajek} Mamajek, E.E.  2005, \apj, 634, 1385

\bibitem[Marley et al.(2002)]{frain} Marley, M.S., Seager, S., Saumon, D., Lodders, K., Ackerman, A.S., Freedman, R.S., \& Fan, X.  2002, \apj, 568, 335

\bibitem[Marley et al.(2007)]{marley} Marley, M.S., Fortney, J.J., Hubickyj, O., Bodenheimer, P., \& Lissauer, J.J.  2007, \aj, 655, 541

\bibitem[Mohanty et al.(2003)]{mohantyvrad} Mohanty, S., Jayawardhana, R.,
\& Barrado y Navascues, D.  2003, \apj, 593, L109

\bibitem[Mohanty et al.(2007)]{mohantydisk} Mohanty, S., Jayawardhana, R., Huelamo, N., \& Mamajek, E.  2007, \apj, 657, 1064

\bibitem[Osten \& Jayawardhana(2006)]{radio} Osten, R.A., \& Jayawardhana, R.  2006, \apj 644, L67

\bibitem[Perryman et al.(1997)]{hipparcos} Perryman, M.A.C., et al.
1997, \aap, 323, L49

\bibitem[Riaz et al.(2006)]{spitzer} Riaz, B., Gizis, J.E., \& Hmiel, A.  2006, \apj, 639, L79

\bibitem[Riaz \& Gizis(2007)]{riaz} Riaz, B., \& Gizis, J.E. 2007, \apj, 661, 354

\bibitem[Scholz \& Jayawardhana(2006)]{broadha} Scholz, A., \& Jayawardhana, R.  2006, \apj, 638, 1056

\bibitem[Skrutskie et a.(2006)]{skrutskie} Skrutskie, M.F., et al. 2006, \aj, 131, 1163 

\bibitem[Song et al.(2006)]{song} Song, I., et al. 2006, \apj, 652, 724

\bibitem[Sterzik et al.(2004)]{gemini} Sterzik, M.F., Pascucci, I., Apai, D., van der Bliek, N,
\& Dullemond, C.P.  2004, \aap, 427, 245

\bibitem[Tinney et al.(2003)]{tinney} Tinney, C.G., Burgasser, A.J., Kirkpatrick, J.D. 2003, \aj, 126, 975

\bibitem[Tsuji(2005)]{tsuji} Tsuji, T.  2005, \apj, 621, 1033

\bibitem[Webb et al.(1999)]{twa} Webb, R.A., et al. 1999, \apj, 512, L63

\bibitem[Whelan et al.(2007)]{outflow} Whelan, E.T., Ray, T.P., Randich, S., Bacciotti, F., Jayawardhana, R., Testi, L., Natta, A., \& Mohanty, S.  2007, \apj, 659, L45

\bibitem[Vrba et al.(2004)]{vrba} Vrba, F.J., et al. 2004, \aj, 127, 2948

\bibitem[Zuckerman et al.(2001)]{betapic} Zuckerman, B., Song, I., Bessell, M.S., \& Webb, R.A.  2001, \apj, 562, L87

\end{thebibliography}
\end{document}